\documentclass[twocolumn,english,pre,showpacs]{revtex4}
\usepackage[T1]{fontenc}
\usepackage[latin1]{inputenc}
\usepackage{graphicx}
\usepackage{amssymb}

\makeatletter

\providecommand{\tabularnewline}{\\}

\usepackage{babel}
\makeatother
\begin{document}

\title{Phenomenological Scale Factors for the Liquid-Vapor Critical Transition
of Pure Fluids}

\author{Y. Garrabos}

\affiliation{Laboratoire des Interactions Moléculaires et des Hautes Pressions
du CNRS, Centre Universitaire Paris-Nord, Avenue J. B. Clément, 93430
Villetaneuse, France.}

\begin{abstract}
We discuss a phenomenological method which allows to determine the
singular asymptotic behaviours for a pure fluid at equilibrium, when
the liquid-gas critical point and the tangent plane to the characteristic
surface of this point are localized experimentally, in the pressure,
density and temperature coordinates.
\end{abstract}


\maketitle
English translation of the published paper in J. Physique (Paris)
\textbf{46} (1985) 281-291.

\section{Introduction}

The renormalization group (RG) theory has contributed to the description
of the critical phenomena observed in systems at equilibrium very
close to their second order transition points where the order parameter
of the transition goes to zero continuously \cite{Cargese1980}. Very
different physical systems can be grouped within the same universality
class \cite{Cargese1980,Kadanoff1970} if they have the same dimension
$d$ of the space, and the same dimension $n$ of the order parameter
(OP) density, respectively. For each given class, the RG theory predicts
the existence of two dimensionless parameters characteristic of each
system belonging to this class \cite{Cargese1980,Baker1976,LeGuillou1981,Bervillier1980,Bagnuls1981}.
This result is a verification of the two-scale factor universality
hypothesis \cite{Stauffer1972}. However, the theory is unable to
provide values for these two parameters which are system-dependent. 

In the present study, we propose a phenomenological determination
of these two characteristic parameters for a pure fluid at equilibrium
very close to its gas-liquid critical point (CP). We demonstrate that
the asymptotic singular behaviour of such near-critical fluid can
be completely defined by four different critical parameters which
are: i) the critical pressure, $p_{c}$; ii) the critical temperature,
$T_{c}$; iii) the critical density, $\rho_{c}$; iv) the common limiting
critical slope, $\gamma_{c}^{'}$, in the $p,T$ diagram, of the saturation
vapor pressure curve and the critical isochore at the critical point.
As a result, the two scale factors characteristic of each pure fluid
can be related to these four quantities.

In order to derive the explicit forms of the two scale factors, we
select the two asymptotic amplitudes, $\xi^{+}$ and $D^{c}$ as being
the two independent (but each one system-dependent) parameters. $\xi^{+}$
is the asymptotic amplitude for the dimensionless correlation length,
$\ell^{*},$ along the critical isochore in the single phase region.
$D^{c}$ is the amplitude for the dimensionless chemical potential
difference, $\Delta\mu_{\bar{p}}^{*}=\mu_{\bar{p}}^{*}-\mu_{\bar{p},c}^{*}$,
along the critical isotherm (here the chemical potential refers to
the normalized thermodynamics per particle). All the pure fluids constitute
a sub-class within the universality class $d=3$ and $n=1$ \cite{Lee1952,Fisher1967},
for which the asymptotic behaviour is then completely determined provided
that the same definitions for dimensionless quantities can be used
for any pure fluid. Therefore, we also propose in this study an explicit
selection of the two macroscopic parameters of energy dimension and
length dimension, respectively, which are necessary to formulate a
description of the critical phenomena observed close to the CP in
terms of the dimensionless variables.

In the next section 2, we recall the main results of the RG theory
for $d=3$ and $n=1$ which are needed for a correct description (in
terms of $\xi^{+}$ and $D^{c}$) of the universal features of the
critical behavior for the pure fluid subclass. In section 3, we unequivocally
define four different quantities characteristic of pure fluids in
terms of $T_{c}$, $p_{c}$, $\rho_{c}$ and $\gamma_{c}^{'}$. Then
we introduce $\left(\beta_{c}\right)^{-1}=k_{B}T_{c}$ and $\alpha_{c}=\left(\frac{k_{B}T_{c}}{p_{c}}\right)^{\frac{1}{d}}$
($k_{B}$ is the Boltzmann's constant), as units of length and energy,
respectively, to made dimensionless all the thermodynamics properties
and correlations functions. The remaining two dimensionless parameters
are $Y_{c}=\gamma_{c}^{'}\frac{T_{c}}{p_{c}}-1$ and $Z_{c}=\frac{p_{c}m_{\bar{p}}}{\rho_{c}k_{B}T_{c}}$
($m_{\bar{p}}$ is the molecular mass of the fluid particle). In section
4, we give relations which enable the unequivocal calculation of the
amplitudes, $\xi^{+}$ and $D^{c}$, knowing $Y_{c}$ and $Z_{c}$.
The agreement with the existing experimental data is very good. Using
these relations, we propose in section 5, a scale dilatation method
of the physical fields which provide a formulation of the asymptotic
universality for the pure fluid subclass equivalent to the one obtained
from the RG methods applied to the complete universality class. For
crossing CP along the thermodynamic path corresponding to the critical
isochore, this formulation uses quantities such as pressure and temperature
which are directly accessible with experiments. We conclude after
a qualitative comparison to the recent results of Bagnuls and Bervillier
\cite{Bagnuls1984a} given in section 6.

\section{Recall of the main results obtained from RG Theory}

The theoretical results will be concerned with systems (noted $S$)
belonging to the universality class, $d=3$ and $n=1$. Our presentation
will be limited to the characterization of the asymptotic singular
behavior of a property, $X_{S}$, which is expected to follow the
non-analytic power-law\begin{equation}
X=X_{S}^{(L)}\left(\tau_{S}^{*}\right)^{-x}\left[1+\mathcal{O}\left(\tau_{S}^{*}\right)\right]\label{X power law; JP Eq. (1)}\end{equation}
when $\tau_{S}^{*}$ goes to zero. All the quantities which appear
in Eq. (\ref{X power law; JP Eq. (1)}) are made dimensionless (see
below). $\tau_{S}^{*}$ is a small finite physical field, associated
to a thermodynamic path denoted by the subscript $\left(L\right)$,
which vanishes asymptotically to CP ($\tau_{S}^{*}=0$ at CP). When
the subscript $\left(L\right)$ denote the \emph{critical isochore}
of $S$, the associated field $\tau_{S}^{*}$ can be thus defined
as the \emph{thermal} field, $\tau_{S}^{*}\equiv\Delta\tau^{*}=\frac{T-T_{c}}{T_{c}}$,
which is the dimensionless temperature distance to CP for fluid systems.
Along this critical isochore, $\left(L\right)$ will be then noted
$+$ when the critical point is approached from the single phase region
$(T>T_{c};\Delta\tau^{*}>0)$ and -
 when the critical point is approached from the two-phase region $(T<T_{c};\Delta\tau^{*}<0)$.
The exponent $x$ is a universal number which depends not on the system,
but on the universality class that the system belongs to. The reduced
asymptotic amplitude, $X_{S}^{(L)}$ is a system dependent quantity.
The term $\left[1+\mathcal{O}\left(\tau_{S}^{*}\right)\right]$ corresponds
to the so-called Wegner expansion \cite{Wegner1972}, which accounts
for the deviation from power-law behaviour with increasing $\tau_{S}^{*}$
due to the confluent corrections to scaling. Such a corrective expansion
will not be taken into consideration in this study \cite{Confluent}.
Similar power law formulations of the singular behaviors of $X_{S}$
can be provided in terms of the vanishing \emph{ordering} field, hereafter
noted $h_{S}^{*}$ (with $h_{S}^{*}=0$ at CP), when the thermodynamic
path to cross CP corresponds to the critical isotherm (with $\left(L\right)\equiv c$
in that case). 

For any system $S$ belonging to the universality class $d=3$ and
$n=1$, noted $\phi_{d=3}^{4}\left(1\right)$ in the following, the
RG numerical results obtained from the non linear treatment of $\phi^{4}$
model in three dimensions \cite{Baker1976,LeGuillou1981,Bervillier1980},
are of sufficient accuracy to verify the two-scale factor universality
hypothesis. Therefore, there are only ten universal relations between
the twelve asymptotic amplitudes as there are only ten scaling laws
related to twelve critical exponents \cite{Bervillier1976}. As a
result, the non-universality of $\phi_{d=3}^{4}\left(1\right)$, can
be adequately characterized by selection of two independent asymptotic
amplitudes. In this present study, we chose the two amplitudes, $\xi^{+}$
and $D^{c}$, defined as follows:

i) along the critical isochore in the single phase region ($\tau_{S}^{*}>0;\, h_{S}^{*}=0;\,\left(L\right)\equiv+$)

\begin{equation}
\ell_{S}=\xi^{+}\left(\tau_{S}^{*}\right)^{-\nu}\label{corr length power law; JP Eq. (2)}\end{equation}

ii) along the critical isotherm ($\tau_{S}^{*}=0;\, h_{S}^{*}\neq0;\,\left(L\right)\equiv c$)

\begin{equation}
h_{S}^{*}=D^{c}\left|m_{S}^{*}\right|^{\delta}\label{ordering field power law; JP Eq. (3)}\end{equation}
where $\nu$ and $\delta$ are the two critical exponents of respective
universal value. $m_{S}^{*}$ is the order parameter (OP) density
for the transition, while $h_{S}^{*}$ is it congugate ordering field.
$\ell_{S}^{*}$ is the correlation length of the OP density fluctuations.
For the case where $S$ is a one-component fluid in the close vicinity
to its CP, (with the fluid subclass noted $VLCP$ in the following),
OP density is proportional to the difference between the \emph{number}
density ($n=\frac{N}{V}=\frac{\rho}{m_{\bar{p}}}$) and the critical
number density ($n_{c}=\frac{\rho_{c}}{m_{\bar{p}}}$), i.e. $m_{S}^{*}\equiv\Delta n^{*}\propto\Delta n=n-n_{c}=\frac{\Delta\rho}{m_{\bar{p}}}=\frac{\rho-\rho_{c}}{m_{\bar{p}}}$
($N$ is the total amount of matter, i.e. the total number of constitutive
particles, and $V$ is the total volume of the system). Its conjugate
\emph{ordering} field is then proportional to the difference between
the chemical potential per particle ($\mu_{\bar{p}}=\mu_{\rho}\times m_{\bar{p}}$)
and the critical chemical potential ($\mu_{\bar{p},c}=\mu_{c}\times m_{\bar{p}}$),
i.e. $h_{S}^{*}\equiv\Delta\mu_{\bar{p}}^{*}\propto\Delta\mu_{\bar{p}}=\mu_{\bar{p}}-\mu_{\bar{p},c}=\Delta\mu\times m_{\bar{p}}=\left(\mu_{\rho}-\mu_{\rho,c}\right)\times m_{\bar{p}}$.
We note that the \emph{mass} density $\rho$ and the chemical potential
per mass unit $\mu_{\rho}$ are the two conjugated density-field variables
when the thermodynamics is normalized per mass unit \cite{Levelt1981}.

\begin{table*}
\begin{tabular}{|cc|c|c|c|c|}
\hline 
\multicolumn{2}{|c|}{a. Universality class}&
Amplitudes&
Amplitude combinations&
Exponents&
Scaling laws\tabularnewline
\hline
\begin{tabular}{c}
Thermodynamics\tabularnewline
\&\tabularnewline
Correlations\tabularnewline
\end{tabular}&
\begin{tabular}{c}
Independent \tabularnewline
quantities\tabularnewline
\end{tabular}&
$\begin{array}{c}
\xi^{+}\\
D^{c}\end{array}$&
&
$\begin{array}{c}
\nu=0.63\\
\delta=4.815\end{array}$&
\tabularnewline
\cline{2-2} \cline{3-3} \cline{4-4} \cline{5-5} \cline{6-6} 
\multicolumn{1}{|c}{}&
Universality&
$\begin{array}{c}
A+\\
\hat{D}\end{array}$&
$\begin{array}{c}
R_{\xi}^{+}=\xi^{+}\left(A^{+}\right)^{\frac{1}{d}}\\
R_{D}=D^{c}\left(\hat{D}\right)^{\frac{\delta+1}{2}}\end{array}$&
$\begin{array}{c}
\alpha\\
\eta\end{array}$&
$\begin{array}{c}
d\nu=2-\alpha\\
\frac{2-\eta}{d}=\frac{\delta-1}{\delta+1}\end{array}$\tabularnewline
\hline 
\multicolumn{1}{|c}{Thermodynamics}&
\begin{tabular}{c}
Independent \tabularnewline
quantities\tabularnewline
\end{tabular}&
$D^{c},\; A^{+}$&
&
$\delta,\;\alpha$&
\tabularnewline
\cline{2-2} \cline{3-3} \cline{4-4} \cline{5-5} \cline{6-6} 
\multicolumn{1}{|c}{}&
Universality&
$\begin{array}{c}
\Gamma^{c}\\
A^{c}\\
\Gamma^{+}\\
B\\
\Gamma^{-}\\
A^{-}\end{array}$&
$\begin{array}{c}
l=\delta\Gamma^{c}\left(D^{c}\right)^{\frac{1}{\delta}}\\
Q_{A^{c}}=A^{c}\left[\delta A^{+}\left(D^{c}\right)^{\frac{1}{\delta}}\right]^{\frac{-\alpha}{2-\alpha}}\\
Q_{\Gamma^{+}}=\left(\Gamma^{+}\right)^{\frac{\delta+1}{\delta}}\left[\left(A^{+}\right)^{\delta-1}\left(D^{c}\right)^{2}\right]^{\frac{1}{\delta}}\\
Q_{B}=B\left(\frac{D^{c}}{A^{+}}\right)^{\frac{1}{\delta+1}}\\
\Gamma^{+}/\Gamma^{-}\\
A^{+}/A^{-}\end{array}$&
$\begin{array}{c}
\gamma^{c}\\
\alpha^{c}\\
\gamma\\
\beta\\
\gamma^{'}\\
\alpha^{'}\end{array}$&
$\begin{array}{c}
\gamma^{c}=\frac{\delta-1}{\delta}\\
\alpha^{c}=\frac{\alpha\left(\delta+1\right)}{2-\alpha}\\
\gamma=\frac{\left(2-\alpha\right)\cdot\left(\delta-1\right)}{\delta+1}\\
\beta=\frac{2-\alpha}{\delta+1}\\
\gamma=\gamma^{'}\\
\alpha=\alpha^{'}\end{array}$\tabularnewline
\hline 
Correlations&
\begin{tabular}{c}
Independent \tabularnewline
quantities\tabularnewline
\end{tabular}&
$\xi^{+},\;\hat{D}$&
&
$\nu,\;\eta$&
\tabularnewline
\cline{2-2} \cline{3-3} \cline{4-4} \cline{5-5} \cline{6-6} 
\multicolumn{1}{|c}{}&
Universality&
$\begin{array}{c}
\xi^{c}\\
\xi^{-}\end{array}$&
$\begin{array}{c}
Q_{\xi^{c}}=\xi^{c}\left(\hat{D}\right)^{\frac{\delta+1}{2d\delta}}\\
\xi^{+}/\xi^{-}\end{array}$&
$\begin{array}{c}
\nu^{c}\\
\nu^{'}\end{array}$&
$\begin{array}{c}
\nu^{c}=2/\left(d+2-\eta\right)\\
\nu=\nu^{'}\end{array}$\tabularnewline
\hline
\end{tabular}

\begin{tabular}{|cc|c|c|c|c|}
\hline 
\multicolumn{6}{|c|}{b. relations to the amplitude combinations defined in Table 1 of reference
\cite{Bervillier1976}}\tabularnewline
\hline 
\multicolumn{2}{|l|}{$\begin{array}{l}
Q_{A^{c}}=\left[R_{A^{c}}/\left(R_{\xi^{+}}\right)^{2d}\right]^{-\alpha/\left(2-\alpha\right)}\\
Q_{\Gamma^{+}}=\left[\left(R_{\chi}\right)^{2}\left(R_{C}\right)^{\delta-1}\right]^{1/\delta}\\
Q_{B}=\left(R_{\chi}/R_{C}\right)^{1/\left(\delta+1\right)}\\
Q_{\xi^{c}}=\left[Q_{2}Q_{3}/\delta\left(R_{D}\right)^{1/\delta}\right]^{2\delta/\left(\delta-1\right)}\end{array}$}&
\multicolumn{2}{c|}{$\left\{ \begin{array}{l}
R_{\xi}=\Gamma^{+}DB^{\delta-1}\\
R_{C}=A^{+}\Gamma^{+}/B^{2}\\
R_{A^{c}}=\left[\left(\xi^{+}\right)^{2}\left(A^{c}\right)^{\nu}\right]^{d}\left(\Gamma^{c}\right)^{\alpha}\end{array}\right.$}&
\multicolumn{2}{c|}{$\left\{ \begin{array}{l}
Q_{1}=\Gamma^{c}\delta\left(B^{\delta-1}\Gamma^{+}\right)^{-1/\delta}\\
Q_{2}=\left(\Gamma^{+}/\Gamma^{-}\right)\left(\xi^{c}/\xi\right)^{2-\eta}\\
Q_{3}=\hat{D}\left(\xi^{+}\right)^{2-\eta}/\Gamma^{+}\end{array}\right.$}\tabularnewline
\hline
\end{tabular}

\caption{a. Equivalence between the ten universal amplitude combinations and
the ten scaling laws connecting the twelve asymptotic amplitudes and
the twelve critical exponents, respectively, for the universality
class $d=3$ and $n=1$. The definitions and notations are those used
in Ref. \cite{Bervillier1976}, except for the four new universal
ratios $Q_{A^{c}}$, $Q_{\Gamma^{+}}$, $Q_{B}$ (which have been
introduced in the {}``Thermodynamics'' part), and $Q_{\xi^{c}}$
(which have been introduced in the {}``Correlations'' part) of this
Table. Of course, they are connected to the other universal ratios
proposed in Ref. \cite{Bervillier1976}, as shown in the lower part
b. In part a, each mixed description and each separated description
of the thermodynamic properties and the correlation functions are
made in terms of the two amplitude-exponent pairs selected as being
independent. }
\end{table*}

In Table 1, we present the results of RG treatment where we can easily
verify that the two numbers, $\xi^{+}$ and $D^{c}$ are sufficient
for complete calculation of all the other asymptotic amplitudes \cite{notations}.
The estimated values of the two independent exponents, $\nu$ and
$\delta$, are $0.63$ and $4.815$, respectively \cite{LeGuillou1981}. 

We emphasize however that this RG formulation of universal features
for any complete class is only the consequence of the basic assumptions
in this theoretical approach \cite{Wilson1971}. Indeed, this approach
assumes analytical relations between the two relevant renormalized
fields, $t^{*}$ and $h^{*}$, of the model, and the corresponding
physical fields, $\tau_{S}^{*}$ and $h_{S}^{*}$, of the system.
In the asymptotic near critical domain, these relations are given
by\begin{equation}
t^{*}=\vartheta\tau_{S}^{*}+\mathcal{O}\left[\left(\tau_{S}^{*}\right)^{2},\tau_{S}^{*}h_{S}^{*}\right]\label{renorm tstar; JP Eq. (4)}\end{equation}

\begin{equation}
h^{*}=\psi h_{S}^{*}+\mathcal{O}\left[\left(h_{S}^{*}\right)^{2},\tau_{S}^{*}h_{S}^{*}\right]\label{renorm hstar; JP Eq. (5)}\end{equation}
where the scale factors $\vartheta$ and $\psi$ are mandatory system-dependent
parameters. The concept of two-scale factor universality lies behind
the direct linearization of equations (\ref{renorm tstar; JP Eq. (4)})
and (\ref{renorm hstar; JP Eq. (5)}) \cite{Wilson1971,Bagnuls1984a},
according to which each system is characterized uniquely by the two
factors, $\vartheta$ and $\psi$, linked to $\tau_{S}^{*}$ and $h_{S}^{*}$,
respectively \cite{Wilson1971,Bagnuls1984a}. It is therefore equivalent
to characterize the non-universality of the fluid, either by two independent
amplitudes, such as $\xi^{+}$ and $D^{c}$, or by the two scale factors,
$\vartheta$ and $\psi$ \cite{Bagnuls1984b}.

\section{Characteristic parameters of VLCP}

The singular thermodynamic properties of a pure fluid are usually
expressed in reduced form using $T_{c}$, $p_{c}$ and $\rho_{c}$
\cite{Sengers1980a}. Such a procedure, however presents difficulties
\cite{Garrabos1982}. The results are often given in terms of the
critical compression factor $Z_{c}$ which is defined by the relation\begin{equation}
Z_{c}=\frac{p_{c}m_{\bar{p}}}{\rho_{c}k_{B}T_{c}}\label{compression factor; JP Eq. (6)}\end{equation}
and which is a characteristic parameter of each pure fluid. Moreover,
the knowledege of the three critical parameters, $T_{c}$, $p_{c}$
and $\rho_{c}$ is not suficient to define the thermodynamic paths
of crossing CP. Indeed, the definition of these lines require not
only to localise the critical point in the phase space, but also to
provide the position of the tangent plane with respect to the phase
surface at this point. To precise that topological aspect, we consider
the phase surface of equation $f\left(p,\rho,T\right)=0$ used by
the experimentalists, which results from the equation of state $p\left(v_{\bar{p}},T\right)=-\left(\frac{\partial a_{\bar{p}}}{\partial v_{\bar{p}}}\right)_{T}$,
where $a_{\bar{p}}\left(T,v_{\bar{p}}\right)$ is the Helmoltz energy
per particle, and $v_{\bar{p}}=\frac{1}{n}=\frac{m_{\bar{p}}}{\rho}$
is the volume per particle (considering here a one-component fluid
at constant amount of matter and the thermodynamic potentials normalized
per particle, such as for example $a_{\bar{p}}\left(T,v_{\bar{p}}\right)=\frac{A\left(T,V,N\right)}{N}$,
where $A\left(T,V,N\right)$ is the total Helmoltz free energy). 

The tangent plane to this phase surface at CP is defined by two orthogonal
lines which pass through CP. One direction is trivial - parallel to
the axis of the normalized extensive variable $v_{\bar{p}}$ - and
is common to all the fluids (reflecting infinite value of their isothermal
susceptibility at CP). On the contrary, the second (finite) direction
is characteristic of each pure fluid. It is defined as the common
limiting slope $\gamma_{c}^{'}$ at the critical point, in the $p;T$
diagram, of the saturation vapor pressure curve $p_{sat}\left(T\right)$
and of the \emph{isochoric} pressure $p\left(T,\rho_{c}\right)$ along
the critical isochore. $\gamma_{c}^{'}$ can be expressed as\begin{equation}
\gamma_{c}^{'}=\frac{\partial}{\partial T}\left[p\left(T,\rho_{c}\right)\right]_{T\rightarrow T_{c}^{+}}=\frac{d}{dT}\left[p_{sat}\left(T\right)\right]_{T\rightarrow T_{c}^{-}}\label{pT direction; JP Eq. (7)}\end{equation}

In view of the thermodynamic approach to CP, there are four macroscopic
critical parameters. We unambiguously define four new quantities given
by the following equations \cite{Garrabos1982}

\begin{equation}
\left(\beta_{c}\right)^{-1}=k_{B}T_{c}\label{energy unit; JP Eq. (8a)}\end{equation}

\begin{equation}
\alpha_{c}=\left(\frac{k_{B}T_{c}}{p_{c}}\right)^{\frac{1}{d}}\label{length unit; JP Eq. (8b)}\end{equation}

\begin{equation}
Z_{c}=\frac{p_{c}m_{\bar{p}}}{\rho_{c}k_{B}T_{c}}\label{isothermal scale factor; JP Eq. (6) and (8c)}\end{equation}

\begin{equation}
Y_{c}=\gamma_{c}^{'}\frac{T_{c}}{p_{c}}-1\label{isochoric scale factor; JP Eq. (8d)}\end{equation}
where $\left(\beta_{c}\right)^{-1}$ and $\alpha_{c}$ are two quantities
evaluated at the CP, with dimensions of energy and length, respectively.
They are sufficient to reduce the physical quantities of fluids into
their dimensionless form \cite{dimensionless}. $Z_{c}$ and $Y_{c}$
are then two dimensionless parameters, characteristics of each fluid. 

The analyses of the thermodynamic analogies between systems belonging
to the same universality class \cite{Lee1952,Fisher1967} are at the
origin of the relations (\ref{isothermal scale factor; JP Eq. (6) and (8c)})
and (\ref{isochoric scale factor; JP Eq. (8d)}). These analogies
establish the symmetry properties of the characteristics surfaces
\cite{Fisher1967}. For the case of pure fluids along the critical
isochore in the single phase domain, the normalized field $\frac{p}{T}$
seems thus the analog of the density of the total Gibbs free energy
for the case of Ising systems belonging to $\phi_{d=3}^{4}\left(1\right)$
(considering then the system maintained at constant volume and the
thermodynamics normalized per volume unit) \cite{LeyKoo1977,Fisher1967b}.
The definition of $\alpha_{c}$, such as a length unit of Eq. (\ref{length unit; JP Eq. (8b)}),
arises from the natural reduction of $\frac{p}{T}$ by the corresponding
critical intensive parameters (which are then non-dependent of the
total volume of the system). The fact that the asymptotic divergence
of the specific heat at constant volume along this path is only due
to the contribution of the singular part of $\frac{p}{T}$, is also
taken into account in establishing the dimensionless form of $Y_{c}$
given by Eq. (\ref{isochoric scale factor; JP Eq. (8d)}) \cite{LeyKoo1977,Moldover1980}.

We emphasize that the dimensionless numbers $Z_{c}$ and $Y_{c}$
are not two of the twelve dimensionless asymptotic amplitudes of $\phi_{d=3}^{4}\left(1\right)$
which are defined in Table 1. 

All the pure fluids should show universal behavior predicted by RG
methods is an introductive hypothesis given in the present work. Consequently,
we admit that $\xi^{+}$ and $D^{c}$ are related to $Z_{c}$ and
$Y_{c}$ by both functional forms given below

\begin{equation}
\xi^{+}=\xi^{+}\left(Y_{c},Z_{c}\right)\label{xiplus-yc-zc closed form; JP Eq. (9a)}\end{equation}

\begin{equation}
D^{c}=D^{c}\left(Y_{c},Z_{c}\right)\label{dc-yc-zc closed form; JP Eq. (9b)}\end{equation}
It stands out that this new formulation of our hypothesis has an important
topological characteristic. The functional relations (\ref{xiplus-yc-zc closed form; JP Eq. (9a)})
and (\ref{dc-yc-zc closed form; JP Eq. (9b)}) assume that all the
information necessary to describe the asymptotic behavior in approaching
the CP is contained in the location of this point at the characteristic
surface and the tangent plane to this surface at this point.

\section{Phenomenological determination of $\xi^{+}$and $D^{c}$}

The selection of the amplitudes, $\xi^{+}$ and $D^{c}$, indicates
two paths of thermodynamic approach which are distinct in their thermodynamic
nature. One path is on the critical isochore (a line for a fluid at
\emph{constant extensivity}) and the other path is on the critical
isotherm (a line for a fluid at \emph{constant} \emph{intensivity}). 

Equations (\ref{xiplus-yc-zc closed form; JP Eq. (9a)}) and (\ref{dc-yc-zc closed form; JP Eq. (9b)})
can be simplified if each asymptotic amplitude is only a function
of one of the factors. $Y_{c}$ should be for the critical isochore
since it is the reduced form of the slope of the potential $\frac{p}{_{k_{B}T}}$
along this path. To establish a connection between $Z_{c}$ and the
critical isotherm, we need to resort to the results of GR which show
that the amplitude $D^{c}$ is related to the amplitude $\widehat{D}$
through the universal ratio $R_{D}=D^{c}\left(\widehat{D}\right)^{\frac{\delta+1}{2}}$
(see Table I). Since $\widehat{D}$ has one divergent quantity at
the \emph{exact critical point} ($t^{*}=0;h^{*}=0$), $Z_{c}$ and
$\widehat{D}$ are the only parameters defined uniquely at the \emph{exact}
\emph{criticality}. Therefore, it is legitimate to associate the $Z_{c}$
specific role to the critical isotherm since the direction of the
tangent plane along this line is also defined by this same parameter,
for every pure fluid. We lay down therefore the importance of these
phenomenological observations to introduce two supplementary hypotheses. 

i) \emph{The dimensionless number $Y_{c}$ is characteristic of only
the singular asymptotic domain along the critical isochore. }

ii) \emph{The dimensionless number $Z_{c}$ is characteristic of only
the singular asymptotic domain along the critical isotherm.}

Thus, the above functional forms (\ref{xiplus-yc-zc closed form; JP Eq. (9a)})
and (\ref{dc-yc-zc closed form; JP Eq. (9b)}) can be modified as
follows:

\begin{equation}
\xi^{+}=\xi^{+}\left(Y_{c}\right)\label{xiplus-yc closed form; JP Eq. (10a)}\end{equation}

\begin{equation}
D^{c}=D^{c}\left(Z_{c}\right)\label{dc-zc closed form; JP Eq. (10b)}\end{equation}
The phenomenological arguements which permit to explicit the above
functional forms will be provided in a more detailled manner in the
following sections. However, for practical (critical) dimensional
analysis, the above functional forms (\ref{xiplus-yc closed form; JP Eq. (10a)})
and (\ref{dc-zc closed form; JP Eq. (10b)}) can be already written
as the following equations

\begin{equation}
\xi^{+}\left(Y_{c}\right)^{\phi}=Y_{G}\label{xiplus-yc closure equation; JP Eq. (11a)}\end{equation}

\begin{equation}
D^{c}\left(Z_{c}\right)^{\zeta}=F_{G}\label{dc-zc closure equation; JP Eq. (11b)}\end{equation}
where we expect that $\phi$, $\zeta$, $Y_{G}$ and $F_{G}$ have
the same values for all pure fluids. Since the experimental values
of $Z_{c}$ and $Y_{c}$ are available for a large number of fluids,
there are six unknown quantities in equations (\ref{xiplus-yc closure equation; JP Eq. (11a)})
and (\ref{dc-zc closure equation; JP Eq. (11b)}).

For the determination of the above unknown parameters $\phi$, $\zeta$,
$Y_{G}$ and $F_{G}$, it is in principle sufficient to know the experimental
values for $\xi^{+}$ and $D^{c}$ for only two fluids with different
$Z_{c}$ and $Y_{c}$ values. We emphasize that the asymptotic nature
of the amplitudes $\xi^{+}$ and $D^{c}$ imposes that experiments
to be performed in the region as close as possible to CP, with high
relative accuracy \cite{Levelt1981}, particulary in order to minimize
the contribution of the corrective terms \cite{Greywall1973} in the
Wegner expansion.

The previous conditions are partly realized in the Franhauffer diffraction
experiments \cite{Hocken1976}, and in some light scattering experiments
\cite{Lunacek1971,Cannell1975,Guttinger1981,Garrabos}, for the three
fluids $Xe$, $SF_{6}$, and $CO_{2}$. However, the amplitude $D^{c}$,
which can only be deduced from the results of the diffraction experiments,
remains determined with a large uncertainty, reflecting the impossibility
to observe directly the critical shape of the critical isotherm for
the chemical potential of a pure fluid \cite{Levelt1976}. 

This is why it would be preferable to find another way which would
enable the calculation of $D^{c}$, avoiding the difficulties mentioned
previously, but maintaining always enable the unequivocal determination
of $\phi$, $\zeta$, $Y_{G}$ and $F_{G}$. Indeed, the values of
the three amplitudes $\xi^{+}$, $\Gamma^{+}$ and $B$, where any
two of which are independent, can be obtained with acceptable precision
from experiments. $\Gamma^{+}$ is the amplitude for the power law
divergence of the susceptibility along the critical isochore in the
single phase domain and $B$ is the amplitude of the power law top-shape
of the vapor-liquid coexistence curve in the two phase domain. We
can thus utilize the theoretical universal value of the amplitude
combination $R_{\chi}$ (see Table I) to obtain $D^{c}$ knowing $\Gamma^{+}$
and $B$. More generally, starting with the experimental values for
any amplitude pair made from the three amplitudes $\xi^{+}$, $\Gamma^{+}$
and $B$, the remaining non-selected amplitude and $D^{c}$ can be
calculated from the following universal amplitude combinations (see
Table I)\begin{equation}
R=R_{\xi}^{+}\left(R_{C}\right)^{-\frac{1}{d}}=\xi^{+}\left(\frac{B^{2}}{\Gamma^{+}}\right)^{\frac{1}{d}}\label{ampli combi R; JP Eq. (12)}\end{equation}

\begin{equation}
R_{\chi}=D^{c}B^{\delta-1}\Gamma^{+}\label{ampli combi Rkhi; JP Eq. (13)}\end{equation}
using their estimated values $R_{\xi}^{+}=0.27$, $R_{C}=0.066$,
(i.e. $R=0.66811$), and $R_{\chi}=1.7$ \cite{Bervillier1980}.

In practice, we chose xenon as the reference fluid for VLCP. The specific
contributions of $Z_{c}$ and $Y_{c}$ become evidenced by the $Xe-SF_{6}$
comparison (where only the contribution of $Y_{c}$ is significant)
and the $Xe-CO_{2}$ comparison (where both the contributions are
mixed). In such conditions, by using all the possible combinations
together with experimental values of the three asymptotic amplitudes,
$\Gamma^{+}$, $B$ and $\xi^{+}$, we obtained the following empirical
values of the constants, $\phi$ and $\zeta$\begin{equation}
\phi=1.55\pm0.15\label{practical phi; JP Eq. (14a)}\end{equation}

\begin{equation}
\zeta=-0.12\pm0.06\label{practical zeta; JP Eq. (14b)}\end{equation}
The uncertainty of the exponent $\zeta$, because of its small value,
reflects two unfavorable conditions for its determination, which are:
i) the small differences in the $Z_{c}$ values for the three selected
fluids; ii) the large relative variation of the calculated $D^{c}$
values resulting from the different choices of the entry pair made
of two independent amplitudes.

As of today, there exists a lot amount of experiments for pure fluids
close to their critical points to increase the precision on the estimated
values of the exponents $\phi$ and $\zeta$ using the method described
above. However, our phenomenological approach implies other constraints
on $\phi$ and $\zeta$ which permit their determination, specially
from scaling arguements. This is because the universality of the sub-class
of fluids is necessarily the one predicted by the RG theory. As a
particular point, the two only independent critical exponents are
$\nu$ and $\delta$, selected in section 2 and defined in Table I.
Consequently, in conformity with the two hypothesis formulated at
the beginning of this section, there are also unequivocal relations
between $\phi$ and $\nu$ along the critical isochore and between
$\zeta$ and $\delta$ along the critical isotherm. Here, we immediatly
provide the explicit forms of these relations\begin{equation}
\phi=\frac{1}{\nu},\, with\,\nu=0.63;\,\phi=1.587\label{scaling phi; JP Eq. (15a)}\end{equation}

\begin{equation}
\phi=\frac{-2}{3\left(\delta+1\right)},\, with\,\delta=4.815;\,\zeta=-0.11465\label{scaling zeta; JP Eq. (15b)}\end{equation}
The above relations will be discussed more in depth in sections 5
and 6 where we introduce the \emph{scale dilatation} of the field
distances to the CP on each of these paths which is the only method
able to provide the Eqs. (\ref{xiplus-yc closure equation; JP Eq. (11a)}),
(\ref{dc-zc closure equation; JP Eq. (11b)}), (\ref{scaling phi; JP Eq. (15a)}),
and (\ref{scaling zeta; JP Eq. (15b)}). In such a basic approach,
we can already pointed out that the \emph{linearization} of the thermal
field relation for example (see Eq. (\ref{renorm tstar; JP Eq. (4)}),
added to the recent predictions of Bagnuls and Bervillier \cite{Bagnuls1984a}
for the asymptotic singular power law divergence of the correlation
length along the critical isochore, give evidence for the relation
$\phi=1/\nu$.

With the values of $\phi$ and $\zeta$ given by Equations (\ref{scaling phi; JP Eq. (15a)})
and (\ref{scaling zeta; JP Eq. (15b)}), we obtain the constants $Y_{G}$
and $F_{G}$ by using the values of the amplitudes $\Gamma^{+}$ and
$B$ for xenon, chosen as one standard critical fluid \cite{xenon,Senger1978,Sengers1980b,amplitudes}\begin{equation}
Y_{G}=0.380\label{yg value; JP Eq. (16a)}\end{equation}

\begin{equation}
F_{G}=0.526\label{zg value; JP Eq. (16b)}\end{equation}

A first time limited justification of the phenomenological equations
is provided in Tables II and III.

\begin{table}
\begin{tabular}{|c|cccc|}
\hline 
&
&
$\xi$&
$\left(10^{-10}m\right)$&
\tabularnewline
&
1&
2&
3&
4\tabularnewline
\hline 
$Ar$&
$1.5$&
$1.61$&
&
\tabularnewline
$Kr$&
$1.61$&
$1.73$&
&
\tabularnewline
$Xe$&
$1.76$&
$1.89$&
$1.85\pm0.1$&
$1.86\pm0.1$\tabularnewline
$CO_{2}$&
$1.46$&
$1.56$&
$1.5\pm0.1$&
$1.50\pm0.1$\tabularnewline
$SF_{6}$&
$1.84$&
$1.96$&
$1.9\pm0.1$&
$1.86\pm0.1$\tabularnewline
$C_{2}H_{6}$&
$1.80$&
$1.84$&
$1.8\pm0.1$&
\tabularnewline
$CClF_{3}$&
$1.82$&
&
$1.95\pm0.15$&
\tabularnewline
$NH_{3}$&
$1.35$&
$1.44$&
&
\tabularnewline
$C_{2}H_{4}$&
$1.73$&
$1.89$&
&
\tabularnewline
$CH_{4}$&
$1.64$&
$1.72$&
&
\tabularnewline
$N_{2}$&
$1.54$&
$1.62$&
&
\tabularnewline
$O_{2}$&
$1.48$&
$1.62$&
&
\tabularnewline
$F_{2}$&
$1.39$&
&
&
\tabularnewline
$H_{2}O$&
$1.20$&
$1.34$&
&
\tabularnewline
$CF_{4}$&
$1.66$&
&
&
\tabularnewline
$p-H_{2}$&
$1.63$&
$1.87$&
&
\tabularnewline
$^{4}He$&
$1.92$&
$2.15$&
&
\tabularnewline
$^{3}He$&
$2.31$&
$2.67$&
&
\tabularnewline
$C_{3}H_{8}$&
$1.89$&
$2.04$&
&
\tabularnewline
\hline
\end{tabular}

\caption{Values in $\textrm{Å}$ of the asymptotic critical amplitude $\xi_{0}^{+}$
of the correlation length along the critical isochore ($T>T_{c}$)
for neinteen pure fluids.}

Column 1 : our determination using equations (\ref{length unit; JP Eq. (8b)}),
(\ref{isochoric scale factor; JP Eq. (8d)}), (\ref{xiplus-yc closure equation; JP Eq. (11a)}),
(\ref{scaling phi; JP Eq. (15a)}), and (\ref{yg value; JP Eq. (16a)}),
with $\xi_{0}^{+}=\alpha_{c}\xi^{+}$.

Column 2 : calculated values in a semi-empirical way by Sengers et
al (see Ref. \cite{Sengers}).

Column 3 : experimental values obtained by light scattering intensity
measurements and turbidity measurements by Garrabos et al in the LIMHP
(see Ref. \cite{Garrabos}).

Column 4 : experimental values reported by Sengers and Moldover in
Ref. \cite{Senger1978} and by Sengers in Ref. \cite{Sengers1980b}
(for original references of the experimental works see refs. \cite{Hocken1976,Lunacek1971,Cannell1975,Guttinger1981}).
\end{table}

In Table II, we compare for about twenty or so fluids, the values
of the amplitude $\xi_{0}^{+}=\alpha_{c}\xi^{+}$obtained by using
equations (\ref{length unit; JP Eq. (8b)}), (\ref{isochoric scale factor; JP Eq. (8d)}),
(\ref{xiplus-yc closure equation; JP Eq. (11a)}), (\ref{scaling phi; JP Eq. (15a)})
and (\ref{yg value; JP Eq. (16a)}) with experimental data \cite{Senger1978,Sengers1980b,Garrabos}
and with their values calculated using a semi-empirical approach \cite{Sengers}.

\begin{table*}
\begin{tabular}{|ccc|ccc|}
\hline 
&
&
&
$Xe$&
$SF_{6}$&
$CO_{2}$\tabularnewline
&
$m_{\bar{p}}$&
$\left(10^{-26}kg\right)$&
$21.803$&
$24.252$&
$7.308$\tabularnewline
\hline 
(0)&
$p_{c}$&
$\left(MPa\right)$&
$5.84$&
$3.76$&
$7.3753$\tabularnewline
&
$\rho_{c}$&
$\left(kg\, m^{-3}\right)$&
$1110$&
$737$&
$467.8$\tabularnewline
&
$T_{c}$&
$\left(K\right)$&
$289.74$&
$318.70$&
$304.14$\tabularnewline
&
$\gamma_{c}^{'}$&
$\left(MPa\, K^{-1}\right)$&
$0.119$&
$0.0835$&
$0.170$\tabularnewline
\hline 
(1)&
$\left(\beta_{c}\right)^{-1}=k_{B}T_{c}$&
$\left(10^{-21}J\right)$&
$4.0$&
$4.4$&
$4.2$\tabularnewline
&
$\alpha_{c}=\left(\frac{k_{B}T_{c}}{p_{c}}\right)^{\frac{1}{d}}$&
$\left(10^{-10}m\right)$&
$8.815$&
$10.54$&
$8.29$\tabularnewline
&
$Z_{c}=\frac{p_{c}m_{\bar{p}}}{\rho_{c}k_{B}T_{c}}$&
&
$0.287$&
$0.281$&
$0.274$\tabularnewline
&
$Y_{c}=\gamma_{c}^{'}\frac{T_{c}}{p_{c}}-1$&
&
$4.9$&
$6.08$&
$6.01$\tabularnewline
\hline 
(2)&
$\xi^{+}=\left(\frac{Y_{G}}{Y_{c}}\right)^{\nu}$&
&
$0.1997$&
$0.1745$&
$0.175$\tabularnewline
&
$D^{c}=\left(\frac{F_{G}}{Z_{c}}\right)^{-\frac{3\left(\delta+1\right)}{2}}$&
&
$5.04\,10^{-3}$&
$4.24\,10^{-3}$&
$3.42\,10^{-3}$\tabularnewline
\hline 
(3)&
$\xi_{0}^{+}=\alpha_{c}\xi^{+}$&
$\left(10^{-10}m\right)$&
$1.76$&
$1.84$&
$1.45$\tabularnewline
(4)&
$\xi_{0}^{+}\left(exp\right)$; Table II; Ref. \cite{Sengers1980b}&
&
$1.86\pm0.10$&
$1.86\pm0.10$&
$1.50\pm0.10$\tabularnewline
&
$\xi_{0}^{+}\left(exp\right)$; Table II; Ref. \cite{Garrabos}&
&
$1.85\pm0.10$&
$1.9\pm0.1$&
$1.5\pm0.1$\tabularnewline
\hline 
(3)&
$D_{LGS}^{c}=\left(Z_{c}\right)^{-\left(\delta+1\right)}D^{c}$&
&
$7.2$&
$6.8$&
$6.3$\tabularnewline
(5)&
$D_{LGS}^{c}\left(min\right)$; see Ref. \cite{Garrabos1982}&
&
$6.13$&
$5.51$&
$5.92$\tabularnewline
&
$D_{LGS}^{c}\left(max\right)$; see Ref. \cite{Garrabos1982}&
&
$7.53$&
$7.10$&
$6.93$\tabularnewline
\hline 
(3)&
$B_{LGS}=Z_{c}\left[\left(\frac{R_{\chi}}{D^{c}}\right)\left(\frac{R}{\xi^{+}}\right)^{d}\right]^{\frac{1}{\delta+1}}$&
&
$1.455$&
$1.575$&
$1.59$\tabularnewline
(5)&
$B_{LGS}\left(exp\right)$; see Ref. \cite{Hocken1976}&
&
$1.48$&
$1.56$&
$1.54$\tabularnewline
&
$B_{LGS}\left(exp\right)$; see Ref. \cite{Senger1978}&
&
$1.42\pm0.06$&
$1.62\pm0.06$&
$1.42\pm0.05$\tabularnewline
\hline
(3)&
$\Gamma_{LGS}^{+}=\left(Z_{c}\right)^{2}\left(\frac{R_{\chi}}{D^{c}}\right)^{\frac{2}{\delta+1}}\left(\frac{R}{\xi^{+}}\right)^{2-\eta}$&
&
$0.0565$&
$0.044$&
$0.046$\tabularnewline
(5)&
$\Gamma_{LGS}^{+}\left(exp\right)$; see Ref. \cite{Senger1978}&
&
$0.058\pm0.002$&
$0.046\pm0.004$&
$0.046\pm0.002$\tabularnewline
\hline
\end{tabular}

\caption{Results for $Xe$, $SF_{6}$, and $CO_{2}$ of mass particle $m_{\bar{p}}$.}

Lines (0) : Generalized critical parameters of the pure fluids.

Lines (1) : Characteristic parameters of the pure fluids defined by
Equations (\ref{energy unit; JP Eq. (8a)}) to (\ref{isochoric scale factor; JP Eq. (8d)}).

Lines (2) : Calculated values of the asymptotic amplitudes $\xi^{+}$
and $D^{c}$ from the characteristics parameters, using Eqs. (\ref{energy unit; JP Eq. (8a)})
to (\ref{zg value; JP Eq. (16b)}).

Lines (3) : Calculated values of the asymptotics amplitudes $\xi_{0}^{+}=\alpha_{c}\xi^{+}$,
$D_{LGS}^{c}$, $B_{LGS}$, and $\Gamma_{LGS}^{+}$ (see Refs. \cite{dimensionless,Levelt1976,amplitudes}).
See the Table I and the text for definitions and estimated values
of the amplitude combinations $R$ and $R_{\chi}$.

Lines (4) : Experimental values of $\xi_{0}^{+}$ following Table
II.

Lines (5) : Values of $D_{LGS}^{c}$, $B_{LGS}$, and $\Gamma_{LGS}^{+}$
obtained by different analyses of the data from the the Fraunhofer
diffraction experiments (see Ref. \cite{Hocken1976}).
\end{table*}

In Table III, we compare for $Xe$, $SF_{6}$, and $CO_{2}$, the
values of $\xi^{+}$, $D^{c}$, $B$, and $\Gamma^{+}$, determined
using our method, with their experimental values.

The results given in these Tables show that our hypotheses, introducing
the two critical parameters $Z_{c}$ and $Y_{c}$ as being the characteristic
numbers for the singular domain of pure fluids, are realistic.

\section{Master singular behaviour of thermodynamic potentials}

We have assumed that the contribution of each characteristic factor
can be separated along two distinct lines on the phase surface $f(P,\rho,T)=0$.
As a result, the asymptotic form of at least one among the (non-analytic)
thermodynamic properties which are defined on each line is also entirely
characterized by the factor it is associated with. In all generality,
this thermodynamic property comes from the singular contribution of
the appropriate thermodynamic potential and acts such as an equation
of state. This is why we reformulate our two hypothesis in consideration
of the critical behaviour of two thermodynamic potentials, the selection
of which were already justified by analogy between pure fluids and
magnetic systems of spins \cite{LeyKoo1977}. 

Hypothesis 1: \emph{When scaled by the factor $Y_{c}$, the asymptotic
behavior of $\Delta p^{*}$ - the difference of the dimensionless
thermodynamic potential $\left(\frac{p}{T}\right)^{*}$ from its critical
value $\left(\frac{p}{T}\right)_{c}^{*}$ at CP - along the critical
isochore, in the single phase region, is the same for all pure fluids.}

Hypothesis 2: \emph{When scaled by the factor $Z_{c}$, the asymptotic
behavior of $\Delta\mu_{\bar{p}}^{*}$ - the difference of the dimensionless
chemical potential $\mu_{\bar{p}}^{*}$ and its critical value $\mu_{\bar{p},c}^{*}$
at CP - along the critical isotherm, is the same for all pure fluids.} 

In order to view the results predicted from hypothesis 1 and 2, we
consider the critical behavior, schematized on one appropriate field-potential
diagram, of two different pure fluids, that are labelled by the subscript
$i$, which means $i=I$ and $i=II$. Crossing CP along the thermodynamic
path $\left(L\right)$, with \{$\left(L\right)\equiv\pm$ for the
critical isochore, $\left(L\right)\equiv c$ for the critical isotherm\},
the thermodynamic potential $G$, with $\{ G=\frac{p}{T},\mu_{\bar{p}}\}$,
is a function of the field $g$, with $\{ g=T,n\}$. Figure 1a illustrates
the distinction on the two curves $G_{i}(g)\mid_{(L)}$ with $\{ i=I,II\}$.
At each critical point $g=g_{c}$, the potential has the value $G=G_{c}.$
Of particular interest is the potential difference $\Delta G=G-G_{c}$,
which is a function of the field distance $\Delta g=g-g_{c}$ measured
along $\left(L\right)$. 

The reduction of the quantities $\Delta G$ and $\Delta g$ can involve
appropriate combination of the energy scale factor $\left(\beta_{c}\right)^{-1}$,
see Eq. (\ref{energy unit; JP Eq. (8a)}), and the length scale factor
$\alpha_{c}$, see Eq. (\ref{length unit; JP Eq. (8b)}). The results
of the dimensionless processing are given in Figure 1b where each
curve is represented by the following functional form\begin{equation}
\Delta G^{*}=\left.f_{i}^{*}\left(\Delta g^{*}\right)\right|_{\left(L\right)}\label{G(g) physical form; JP Eq. (17)}\end{equation}

Following hypothesis 1 and 2, we can now transform these curves into
a master curve, the functional equation of which is given by\begin{equation}
\mathcal{G}_{\Pi}^{*}\left(\Delta G^{*}\right)=\mathcal{F}_{\left(L\right)}^{*}\left[\mathcal{D}_{\Pi}^{*}\left(\Delta g^{*}\right)\right]\label{G(g) master functional form; JP Eq. (18)}\end{equation}
where $\mathcal{F}_{(L)}^{*}$is a function which does not depend
on the pure fluid under consideration but only on the thermodynamic
path to cross CP. On the other hand, the new dimensionless functions
$\mathcal{G}_{\Pi}^{*}\left(\Delta G^{*}\right)$ and $\mathcal{D}_{\Pi}^{*}\left(\Delta g^{*}\right)$
which appear on equation (\ref{G(g) master functional form; JP Eq. (18)}),
depend on $\Delta G^{*}$ and $\Delta g^{*}$ through the dimensionless
number $\Pi$ which is characteristic of $\left(L\right)$, with $\Pi=Y_{c}$
for the critical isochore $\left(L\right)\equiv\pm$, and $\Pi=Z_{c}$
for the critical isotherm $\left(L\right)\equiv c$. For explaining
this dependency, we resort to the general view of the scale dilatation
of the physical field \cite{Toulouse1975}, which is characterized
here by the scale factor $\Pi$. The quantities $\Delta G$ and $\Delta g$
transform with dilatation to the following quantities\begin{equation}
\mathcal{G}_{\Pi}^{*}=\left.\Pi^{E}\Delta G^{*}\right|_{\left(L\right)}\label{potential dilatation; JP Eq. (19)}\end{equation}

\begin{equation}
\mathcal{D}_{\Pi}^{*}=\left.\Pi^{e}\Delta g^{*}\right|_{\left(L\right)}\label{field dilatation; JP Eq. (20)}\end{equation}
where $E_{(L)}$ and $e_{(L)}$ are the same values for all the pure
fluids. In such conditions, the master curve along $\left(L\right)$
described by equation (\ref{G(g) master functional form; JP Eq. (18)})
is given in Figure 1c, where the renormalized variables $\mathcal{G}_{\Pi}^{*}$
and $\mathcal{D}_{\Pi}^{*}$ are obtained by equations (\ref{potential dilatation; JP Eq. (19)})
and (\ref{field dilatation; JP Eq. (20)}).

For a better representation of the scale dilatation defined on each
path, we derive now the equations (\ref{G(g) physical form; JP Eq. (17)})
to (\ref{field dilatation; JP Eq. (20)}), and the relations (\ref{xiplus-yc closure equation; JP Eq. (11a)})
and (\ref{dc-zc closure equation; JP Eq. (11b)}) introduced in the
previous section. In these next developments, we will consider only
the non-analytic asymptotic contributions of the different functions
\cite{Confluent}.

\begin{figure}
\includegraphics{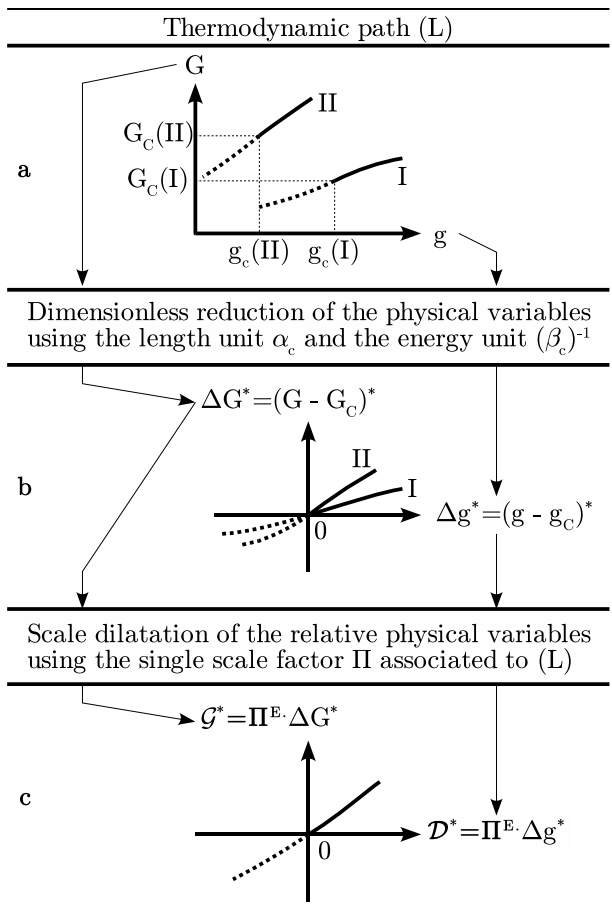}

\caption{a) Schematic asymptotic behaviours along the critical path $\left(L\right)$
of the thermodynamic potential $G$ as a function of the field $g$,
for pure fluids I and II.}

b) Schematic asymptotic behaviours of the dimensionless potential
difference $\Delta G^{*}$ versus the dimensionless field distance
$\Delta g^{*}$ to the critical point. The method to make dimensionless
the potential $G$ and the field $g$, involves separately either
the length scale factor $\alpha_{c}$ {[}see Eq. (\ref{length unit; JP Eq. (8b)}){]}
or the energy one $\left(\beta_{c}\right)^{-1}$ {[}see Eq. (\ref{energy unit; JP Eq. (8a)}){]}.

c) Master asymptotic behaviour {[}see Eq. (\ref{G(g) master functional form; JP Eq. (18)}){]}
along $\left(L\right)$ of the renormalized potential $\mathcal{G}^{*}$
as a function of the renormalized field $\mathcal{D}^{*}$, for every
fluids. This behaviour is obtained using scale dilatations {[}see
Eqs. (\ref{potential dilatation; JP Eq. (19)}) and (\ref{field dilatation; JP Eq. (20)}){]}
of axes in figure 1b, which take into account the scaling factor $\Pi$
related to $\left(L\right)$. $\Pi$ is a characteristic parameter
of each pure fluid.
\end{figure}

\subsection{Scale dilatation along the critical isotherm}

The master scaling form of the renormalized chemical potential, $\mathcal{H}^{*}$
at the critical isotherm is given by \cite{LeyKoo1977,Sengers1980a}\begin{equation}
\mathcal{H}^{*}=\mathcal{Z}_{H}\left|\mathcal{M}^{*}\right|^{\delta}\label{H(m) master form crit isotherm; JP Eq. (21)}\end{equation}
where $\mathcal{M}^{*}$ is the renormalised order parameter ($\mathcal{H}^{*}$
and $\mathcal{M}^{*}$ are two dual variables with respect to one
appropriate renormalized thermodynamic potential). $\mathcal{Z}_{H}$
is a number which has the same value for all pure fluids. In accordance
with equations (\ref{potential dilatation; JP Eq. (19)}) and (\ref{field dilatation; JP Eq. (20)}),
we define the following quantities renormalized along the critical
isotherm as\begin{equation}
\mathcal{H}^{*}=\left(Z_{c}\right)^{V}\Delta\mu_{\bar{p}}^{*}\label{H-mu dilatation; JP Eq. (22)}\end{equation}

\begin{equation}
\mathcal{M}^{*}=\left(Z_{c}\right)^{v}\Delta n^{*}\label{m-rho dilatation; JP Eq. (23)}\end{equation}
The asymptotic amplitude, $D^{c}$, is related to the reduced physical
quantities, $\Delta\mu_{\bar{p}}^{*}$ and $\Delta n^{*}$, by the
following expression\begin{equation}
\Delta\mu_{\bar{p}}^{*}=D^{c}\left|\Delta n^{*}\right|^{\delta}\label{mu-rho phys form crit isotherm; JP Eq. (24)}\end{equation}
As a result, we obtain the relation\begin{equation}
D^{c}=\mathcal{Z}_{H}\left(Z_{c}\right)^{v\delta-V}\label{dc amplitude; JP Eq. (25)}\end{equation}
which assumes the following form\begin{equation}
\left(D^{c}\right)^{\frac{1}{V-v\delta}}Z_{c}=\left(\mathcal{Z}_{H}\right)^{\frac{1}{V-v\delta}}\label{dc-zc master ampli combi; JP Eq. (26)}\end{equation}
equivalent to equation (\ref{dc-zc closure equation; JP Eq. (11b)}).

An examination of equation (\ref{dc-zc master ampli combi; JP Eq. (26)})
shows that the values of $V$ and $v$ are linked to $\delta$ through
equations (\ref{dc-zc closure equation; JP Eq. (11b)}) and (\ref{scaling zeta; JP Eq. (15b)}).
The renormalized variables defined by equations (\ref{H-mu dilatation; JP Eq. (22)})
and (\ref{m-rho dilatation; JP Eq. (23)}) are related by equation
(\ref{H(m) master form crit isotherm; JP Eq. (21)}). Renormalization
by only one variable is sufficient to characterize the scale dilatation
along the critical isotherm. The two new numbers introduced in this
section ($\mathcal{Z}_{H}$ and one exponent $V$ or $v$) are linked
to $F_{G}$ and $\zeta$ introduced previously through the relation
(\ref{dc-zc closure equation; JP Eq. (11b)}). The fact that we have
not been able to determine the value of the exponent $\zeta$ with
certainty in section 4, or the two equivalent exponents $V$ and $v$
in this section, prevents us going further with our phenomenological
approach (due to the fact that all the dimensionless quantities can
be always defined at an unknown power of $Z_{c}$).

\subsection{Scale dilatation along the critical isochore}

When approaching CP along the critical isochore in the single phase
domain, it is possible to quantify the scale dilatation associated
to this path. The reduced differences of physical quantities are expressed
in terms of pressure and temperature in the following manners\begin{equation}
\Delta p^{*}=\frac{p}{T}\frac{T_{c}}{p_{c}}-1\label{grand potential field; JP Eq. (27)}\end{equation}

\begin{equation}
\Delta\tau^{*}\equiv\tau_{S}=\frac{T}{T_{c}}-1\label{thermal field; JP Eq. (28)}\end{equation}
In accordance with our hypothesis 1 and the adopted notation, the
renormalized variables are defined by the following relations\begin{equation}
\mathcal{P}^{*}=\left(Y_{c}\right)^{W}\Delta p^{*}\label{grand pot dilatation; JP Eq. (29)}\end{equation}

\begin{equation}
\mathcal{T}^{*}=\left(Y_{c}\right)^{w}\Delta\tau^{*}\label{thermal field dilatation; JP Eq. (30)}\end{equation}
For $\phi_{d=3}^{4}\left(1\right)$, the non-analytic asymptotic term
of the thermodynamic potential $\mathcal{P}^{*}$ defined along the
critical isochore varies as $\left(\mathcal{T}^{*}\right)^{2-\alpha}$
\cite{LeyKoo1977,Moldover1980}. Particularly for the pure fluid subclass,
experiments predominantly show that $\Delta p^{*}$ has a term linear
in $\Delta\tau^{*}$, which justifies the addition of one regular
contribution, $\left(\propto\mathcal{T}^{*}\right)$, to $\mathcal{P}^{*}$.
Considering the dominant regular and non-analytic terms, the master
scaling form of the potential $\mathcal{P}^{*}$ should be given by\begin{equation}
\mathcal{P}^{*}=a_{1}\mathcal{T}^{*}+\mathcal{Z}_{P}^{+}\left(\mathcal{T}^{*}\right)^{2-\alpha}+...\label{grand potential master form; JP Eq. (31)}\end{equation}
where $a_{1}$ and $\mathcal{Z}_{P}^{+}$ take the same values for
all pure fluids. Taking into account the definition of the characteristic
parameter $Y_{c}$, the selection of values $W=0$ and $w=1$ for
equations (\ref{grand pot dilatation; JP Eq. (29)}) and (\ref{thermal field dilatation; JP Eq. (30)})
implies that $a_{1}=1$. Accordingly, the renormalized variables are
properly defined by the following equations\begin{equation}
\mathcal{P}^{*}\equiv\Delta p^{*}\label{dilated grand potential; JP Eq. (32)}\end{equation}

\begin{equation}
\mathcal{T}^{*}=Y_{c}\Delta\tau^{*}\label{dilated thermal field; JP Eq. (33)}\end{equation}
where $Y_{c}$ plays the role of the multiplicative factor for the
dimensionless thermal distance generally used in experiments. 

With these particular choices, the form of equation (\ref{dc-zc closure equation; JP Eq. (11b)})
described in the previous section is now exact. In fact, the singular
behavior of the renormalized heat capacity $\mathcal{C}^{*}$ is proportional
to the second derivative of the potential considered above. Correspondingly,
the asymptotic amplitude $A^{+}$ of the dimensionless singular part
$\Delta c_{V}^{*}$ of the physical heat capacity at constant volume,
is defined as follows\begin{equation}
\Delta c_{V}^{*}=\frac{A^{+}}{\alpha}\left(\Delta\tau^{*}\right)^{-\alpha}\left[1+...\right]\label{physical specific heat; JP Eq. (34)}\end{equation}
As a result, there is a relation between $A^{+}$ and $Y_{c}$ given
by\begin{equation}
A^{+}=\mathcal{Z}_{A}^{+}\left(Y_{c}\right)^{2-\alpha}\label{aplus ampli; JP Eq. (35)}\end{equation}
where $\mathcal{Z}_{A}^{+}$ is a unique value for every pure fluid
($\mathcal{Z}_{A}^{+}$ depends on the unique value $\mathcal{Z}_{P}^{+}$
and the universal value of $\alpha$, due to the appropriate definition
of the renormalized heat capacity).

As the RG theory predicts the existence of the universal ratio, $R_{\xi}^{+}$
(Table 1), we can deduce the following relation from equation (\ref{aplus ampli; JP Eq. (35)})\begin{equation}
\left(\xi\right)^{\frac{d}{2-\alpha}}Y_{c}=\left(R_{\xi}\right)^{\frac{d}{2-\alpha}}\left(\mathcal{Z}_{A}^{+}\right)^{\frac{1}{2-\alpha}}\label{xi-zc master ampli combi; JP Eq. (36)}\end{equation}
Equation (\ref{xi-zc master ampli combi; JP Eq. (36)}) is equivalent
(from a constant power exponent) to equation (\ref{xiplus-yc closure equation; JP Eq. (11a)})
and is to do with the scaling law of Josephon between the critical
exponents $\alpha$ and $\nu$ (Table I).

The formulation and the steps folllowed in this section for derivation
of equation (\ref{xiplus-yc closure equation; JP Eq. (11a)}) and
(\ref{dc-zc closure equation; JP Eq. (11b)}) characterize the non-universal
transformations of physical fields. We anticipate that with these
transformations, singular behavior of the properties of pure fluids
on each of the thermodynamic paths considered here will be identical.
These phenomenological results are therefore comparable to the predictions
of Bagnuls and Bervillier.

\section{Comparison with the theoretical results obtained by Bagnuls and Bervillier
\cite{Bagnuls1984a}}

Along the critical isochore of pure fluids, we propose the scale dilatation
with a phenomenological definition\begin{equation}
\mathcal{T}^{*}=Y_{c}\tau_{S}^{*}\label{master thermal field linearization; JP recall of Eq. (33)}\end{equation}
which is formally equivalent to the one needed to applied RG thechniques\begin{equation}
t^{*}=\vartheta\tau_{S}^{*}\label{RG thermal field linearization; JP recall of Eq. (4)}\end{equation}
Following the work of Bagnuls and Bervillier, the singular behavior
of the reduced correlation length depends uniquely on the scaled variable
$t^{*}$. The asymptotic term of this behavior is given by\begin{equation}
\ell^{*}\left(t^{*}\right)=\mathcal{\mathbb{Z}}_{\xi}^{+}\left(t^{*}\right)^{-\nu}\left[1+\mathcal{O}\left(t^{*}\right)\right]\label{MR correlation lenth; JP Eq. (37)}\end{equation}
where $\mathcal{\mathbb{Z}}_{\xi}^{+}$ is a {}``universal'' constant.
Using equations (\ref{corr length power law; JP Eq. (2)}) and (\ref{MR correlation lenth; JP Eq. (37)})
and taking into account that $t^{*}$ is given by equation (\ref{renorm tstar; JP Eq. (4)}),
the following relation can be derived\begin{equation}
\xi^{+}=\mathcal{\mathbb{Z}}_{\xi}^{+}\vartheta^{-\nu}\label{MR ksi plus ampli; JP Eq. (38)}\end{equation}
Equation (\ref{MR ksi plus ampli; JP Eq. (38)}) is comparable to
our phenomenological equation (\ref{xiplus-yc closure equation; JP Eq. (11a)})
which has the form given by\begin{equation}
\xi^{+}=\mathcal{Z}_{\xi}^{+}\left(Y_{c}\right)^{-\frac{1}{\phi}}\label{master ksi plus ampli; JP Eq. (39)}\end{equation}
where the relation $\phi=1/v$ is evident between the exponents. On
the other hand, the value of the phenomenological constant given by\begin{equation}
\mathcal{Z}_{\xi}^{+}=\left(Y_{G}\right)^{\nu}\label{master Zksi ampli; JP Eq. (40)}\end{equation}
equals $\mathcal{Z}_{\xi}^{+}=0.54$, as calculated from the $Y_{G}$
value given by equation (\ref{yg value; JP Eq. (16a)}), and is different
from the theoretical value of $\mathcal{\mathbb{Z}}_{\xi}^{+}=0.48$,
provided by the numerical treatment of $\phi_{d=3}^{4}\left(1\right)$.
This difference is essentially due to the non-universality of the
reduction process of the physical quantities. As a matter of fact,
equation (\ref{MR correlation lenth; JP Eq. (37)}) contains the non-universality
of the model, the origin of which is not easy to explain. Nevertheless,
it is easy to conceive that both non-universality and the basic hypothesis
of RG theory are partly taken into account in the dimensionless reduction
of the physical variables. This point is emphasized by Bagnuls and
Bervillier \cite{Bagnuls1984a} and discussed for the case where the
fluid is xenon \cite{Bagnuls1984b}. 

In equation (\ref{master ksi plus ampli; JP Eq. (39)}), we separated
the non-universality of the pure fluid into two terms which are: i)
The dimensionless compression factor, $Z_{c}$ which is associated
with the two chosen units for expressing the variables in reduced
form ; ii) The non-universal number $Y_{c}$ which is the scale factor
along the critical isochore. Our approach assumes therefore that $\mathcal{Z}_{\xi}^{+}$
keeps the same numerical value for all pure fluids. Due to a lack
of a comparison between the results from the model and the ones from
the fluids (specially on the dimensional length scale proper to the
model), there is no evident reason for $\mathcal{Z}_{\xi}^{+}$ to
be a universal number proper to $\phi_{d=3}^{4}\left(1\right)$, and
therefore equal to the $\mathcal{\mathbb{Z}}_{\xi}^{+}$ value found
in Ref. \cite{Bagnuls1984a}.

\section{Conclusions}

We assumed that the asymptotic singular behavior of a pure fluid at
equilibrium close the gas-liquid critical point can be entirely characterized
by four macroscopic parameters. These parameters describe the location
of the critical point on the characteristic surface and the tangent
plane to this surface at this point. Two of these parameters, $\left(\beta_{c}\right)^{-1}=k_{B}T_{c}$
and $\alpha_{c}=\left(\frac{k_{B}T_{c}}{p_{c}}\right)^{\frac{1}{d}}$,
are parameters which are evaluated at CP and have units of energy
and length, respectively. The remaining two parameters are $Y_{c}=\gamma_{c}^{'}\frac{T_{c}}{p_{c}}-1$
and $Z_{c}=\frac{P_{c}m_{\bar{p}}}{\rho_{c}k_{B}T_{c}}$, both of
which are dimensionless numbers. 

In accordance with the two-scale factor universality predicted by
the RG theory, we proposed two relations between the two asymptotic
amplitudes $\xi^{+}$ and $D^{c}$ and the parameters, $Y_{c}$ and
$Z_{c}$. We thus define a new system dependent set made of 12 theoretical
asymptotic amplitudes for $\phi_{d=3}^{4}\left(1\right)$ and two
experimentally accessible parameters, $Y_{c}$ and $Z_{c}$. As a
consequence, the critical singular behavior of each pure fluid which
is defined in RG approach by the selection of two amplitudes is defined
in our phenomenological approach by the four parameters, $\beta_{c}$,
$\alpha_{c}$, $Y_{c}$ and $Z_{c}$. 

We moved forward with two hypothesis for definition of the the singular
behavior of the thermodynamic potential along the critical isochore
and the critical isotherm. As part of these hypothesis, we proposed
a scale factor transformation of the reduced physical fields as functions
of the distance to the critical point along these paths. We obtained
simple relationships between the two asymptotic amplitudes taken as
the system dependent parameters in the RG theory and the characteristic
experimental parameters, $Y_{c}$ and $Z_{c}$. Comparison with the
results of the GR approach shows that the proposed method of scale
factor transformation of physical fields is equivalent to the one
in the theoretical approach. 

We are now working on testing the universality associated with the
contribution of the first term of the non-analytic correction to the
asymptotic behavior \cite{Bagnuls1984a}. 

\begin{acknowledgments}
We thank R. Tufeu and B. Le Neindre for their help and encouragement
throughout this study. We also thank C. Bervillier and C. Bagnuls
for their comments and enriching explanations of the recent developments
of the RG treatment of the $\phi^{4}$ model in three dimensions.

\textbf{Note added 15 December 2005}

To conform with the usual notation of the dimensionless energy such
as $\beta E$ with $\beta=\frac{1}{k_{B}T}$, the present definition
of the energy scale factor corresponds to the inverse of the energy
scale factor defined in the initial french version of this paper.
In addition, the present length scale notation $\alpha_{c}=\frac{k_{B}T_{c}}{p_{c}}$
replaces the initial one $a=\frac{k_{B}T_{c}}{p_{c}}$ (to avoid confuse
situation with the Helmholtz free energy density $a\left(T,n\right)=\frac{A\left(T,V,N\right)}{V}$).
The chemical potential per particle $\mu_{\bar{p}}$ is here labellized
with the decorated subcript $\bar{p}$, to make explicit distinctions
with the chemical potential per mass unit, noted $\mu_{\rho}$, and
with the pressure, noted $p$. The dimensionless leading amplitudes
of the susceptibility, noted $C^{+}$, $C^{-}$ and $C^{c}$ in the
initial french version, are presently noted $\Gamma^{+}$, $\Gamma^{-}$
and $\Gamma^{c}$ (respectively), to conform with actual standard
notations of fluid singularities. In the Table I, a printing power
index mistake was corrected in the definition of the universal combination
$Q_{\Gamma^{+}}$ which now reads $Q_{\Gamma^{+}}=\left(\Gamma^{+}\right)^{\frac{\delta+1}{\delta}}\left[\left(A^{+}\right)^{\delta-1}\left(D^{c}\right)^{2}\right]^{\frac{1}{\delta}}$.
Some font modifications between english and french versions are consistent
with actual fonts used in recent related papers. 
\end{acknowledgments}


\begin{thebibliography}{10}
\bibitem{Confluent}To neglect the contributions related to the confluent corrections
to scaling, due to the probable existence of non-relevant fields,
is without any consequence for the present approach, which is only
based on the needed existence of two relevant scaling fields asymptotically
close to CP.
\bibitem{notations}The notations and definitions of the physical properties which are
not directely used in the present work are those of the Ref. \cite{Bervillier1976}.
\bibitem{amplitudes}The selected experimental data are $B_{LGS}\left(Xe\right)=1.455$
and $\Gamma_{LGS}^{+}\left(Xe\right)=0.0565$, respectively. The relations
which give the corresponding values of $B$ and $\Gamma^{+}$ are
given in Ref. \cite{dimensionless}.
\end{thebibliography}
\end{document}